\documentclass[]{pasj01}

\usepackage{color}
\usepackage{multirow}

\begin{document} 

\title{RCW~36 in the Vela Molecular Ridge:\\Evidence for a high-mass star cluster formation triggered by Cloud-Cloud Collision}

\author{Hidetoshi \textsc{SANO}\altaffilmark{1,2}}%
\author{Rei \textsc{ENOKIYA}\altaffilmark{2}}%
\author{Katsuhiro \textsc{HAYASHI}\altaffilmark{2}}%
\author{Mitsuyoshi \textsc{YAMAGISHI}\altaffilmark{3}}%
\author{Shun \textsc{SAEKI}\altaffilmark{2}}%
\author{Kazuki \textsc{OKAWA}\altaffilmark{2}}%
\author{Kisetsu \textsc{TSUGE}\altaffilmark{2}}%
\author{Daichi \textsc{TSUTSUMI}\altaffilmark{2}}%
\author{Mikito \textsc{KOHNO}\altaffilmark{2}}%
\author{Yusuke \textsc{HATTORI}\altaffilmark{2}}%
\author{Satoshi \textsc{YOSHIIKE}\altaffilmark{2}}%
\author{Shinji \textsc{FUJITA}\altaffilmark{2}}%
\author{Atsushi \textsc{NISHIMURA}\altaffilmark{2}}%
\author{Akio \textsc{OHAMA}\altaffilmark{2}}%
\author{Kengo \textsc{TACHIHARA}\altaffilmark{2}}%
\author{Kazufumi \textsc{TORII}\altaffilmark{4}}%
\author{Yutaka \textsc{HASEGAWA}\altaffilmark{5}}%
\author{Kimihiro \textsc{KIMURA}\altaffilmark{5}}%
\author{Hideo \textsc{OGAWA}\altaffilmark{5}}%
\author{Graeme F. \textsc{WONG}\altaffilmark{6,7}}%
\author{Catherine \textsc{BRAIDING}\altaffilmark{7}}%
\author{Gavin \textsc{ROWELL}\altaffilmark{8}}%
\author{Michael G. BURTON,\altaffilmark{9}}%
\author{Yasuo \textsc{FUKUI}\altaffilmark{1,2}}%
\email{sano@a.phys.nagoya-u.ac.jp}
\altaffiltext{1}{Institute for Advanced Research, Nagoya University, Furo-cho, Chikusa-ku, Nagoya 464-8601, Japan}
\altaffiltext{2}{Department of Physics, Nagoya University, Furo-cho, Chikusa-ku, Nagoya 464-8601, Japan}
\altaffiltext{3}{Institute of Space and Astronautical Science, Japan Aerospace Exploration Agency, Chuo-ku, Sagamihara 252-5210, Japan}
\altaffiltext{4}{Nobeyama Radio Observatory, Minamimaki-mura, Minamisaku-gun, Nagano 384-1305, Japan}
\altaffiltext{5}{Department of Physical Science, Graduate School of Science, Osaka Prefecture University, 1-1 Gakuen-cho, Naka-ku, Sakai, Osaka 599-8531, Japan}
\altaffiltext{6}{Western Sydney University, Locked Bag 1797, Penrith, 2751 NSW, Australia}
\altaffiltext{7}{School of Physics, University of New South Wales, Sydney, NSW 2052, Australia}
\altaffiltext{8}{School of Physical Sciences, University of Adelaide, North Terrace, Adelaide, SA 5005, Australia}
\altaffiltext{9}{Armagh Observatory and Planetarium, College Hill, Armagh BT61 9DG, Northern Ireland, UK}


\maketitle

\begin{abstract}
A collision between two molecular clouds is one possible candidate for high-mass star formation. The H{\sc ii} region RCW~36, located in the Vela molecular ridge, contains a young star cluster {($\sim$1 Myr-old) and} two O-type stars. We present new CO observations of RCW~36 with NANTEN2, Mopra, and ASTE using $^{12}$CO($J$ = 1--0, 2--1, 3--2) and $^{13}$CO($J$ = 2--1) emission lines. We have discovered two molecular clouds lying at the velocities $V_\mathrm{LSR} \sim$5.5 and 9 km s$^{-1}$. Both clouds are likely to be physically associated with the star cluster, as verified by the good spatial correspondence among the two clouds, infrared filaments, and the star cluster. We also found a high intensity ratio of $\sim$0.6--1.2 for CO $J$ = 3--2 / 1--0 toward both clouds, indicating that the gas temperature has been increased due to heating by the O-type stars. We propose that {the O-type stars in RCW~36} were formed by a collision between the two clouds, with a relative velocity separation of 5 km s$^{-1}$. The complementary spatial distributions and the velocity separation of the two clouds are in good agreement with observational signatures expected for O-type star formation triggered by a cloud-cloud collision. We also found a displacement between the complementary spatial distributions of the two clouds, which we estimate to be {0.3} pc assuming the collision angle to be 45$^{\circ}$ relative to the line-of-sight. We estimate the collision timescale to be {$\sim$10$^5$ yr. It is probable that the cluster age by \authorcite{2013A&A...558A.102E} (\yearcite{2013A&A...558A.102E}, A\&A, 558, A102) is dominated by the low-mass members which were not formed under the triggering by cloud-cloud collision, and that the O-type stars in the center of the cluster are explained by the collisional triggering independently from the low-mass star formation.}
\end{abstract}

\section{Introduction} \label{sec:intro}
High-mass stars have a significant influence on the dynamics of the interstellar medium via ultraviolet radiation, stellar winds, and supernova explosions. However, the formation mechanism of high-mass stars remains elusive (e.g., \cite{2007ARA&A..45..481Z}).

Recent observational studies have strongly suggested that high-mass stars associated with H{\sc ii} regions were formed by cloud-cloud collisions (e.g., \cite{2009ApJ...696L.115F, 2011ApJ...738...46T,2015ApJ...807L...4F}). These authors found that two molecular clouds associated with an H{\sc ii} region have complementary spatial distributions and typical velocity separations of $\sim$10--20 km s$^{-1}$ (c.f., \cite{2017arXiv170104669F}). This velocity separation is too large to be generated by the stellar winds from the high-mass stars. Both clouds are also associated with high-mass stars (e.g., \cite{2009ApJ...696L.115F}). In addition, the two molecular clouds are generally connected to each other in velocity space (hereafter termed the ``bridging feature''). Moreover, a high intensity ratio of CO $J$ = 3--2 / 1--0 has been reported toward the two clouds, supporting their physical association with the H{\sc ii} region (e.g., \cite{2015ApJ...806....7T, 2016ApJ...820...26F}). The authors concluded that high-mass star formation is  triggered by the strong compression produced during the cloud-cloud collision. Theoretical {study} also supports the idea that the cloud-cloud collision increases the gas density and the effective sound speed in the shock-compressed layer (e.g., \cite{1992PASJ...44..203H,2010MNRAS.405.1431A,2013ApJ...774L..31I,2014ApJ...792...63T}).

As of early 2017, firm observational evidence for cloud-cloud collision triggering high-mass star formation had been obtained from only ten sources: RCW~38, NGC~3603, Westerlund~2, [DBS2003]~179, the Orion~nebula~cluster, M20, RCW~120, N159W-South, N159E-Papillon, and G018.15$-$00.28 (\cite{2014ApJ...780...36F,2015ApJ...807L...4F,2016ApJ...820...26F,2009ApJ...696L.115F,2010ApJ...709..975O,2011ApJ...738...46T,2015ApJ...806....7T,2010A&A...510A..32M,kuwahara_prep,2017ApJ...835..108S,2016MNRAS.458.2813V}). To understand better the formation mechanism of high-mass stars, we need to have a larger sample of cloud-cloud collisions.

\begin{figure}
\begin{center}
\includegraphics[width=\linewidth]{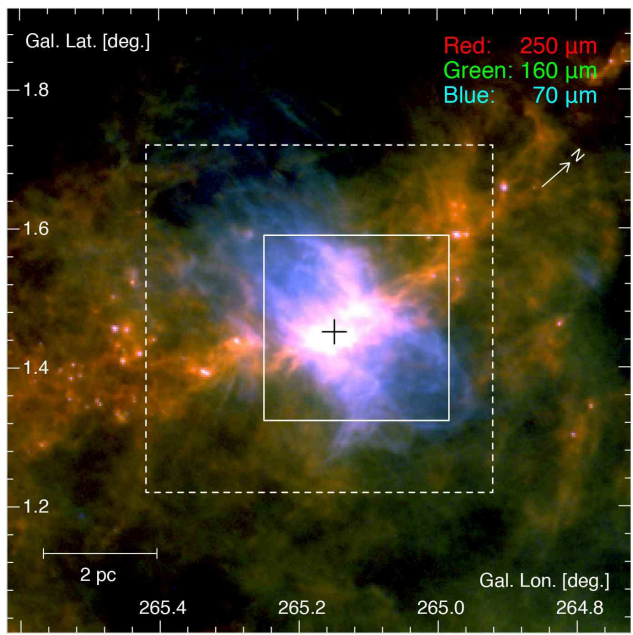}
\end{center}
\caption{RGB image of the H{\sc ii} region RCW~36 obtained with $Herschel$ \citep{2011A&A...533A..94H}. The red, green, and blue colors represent 250-$\mu$m, 160-$\mu$m, and 70-$\mu$m, respectively. Black cross indicates the position of two O-type stars \citep{2013A&A...558A.102E}. Solid and dashed boxes represent the observed regions with the ASTE $^{12}$CO($J$ = 3--2) and NANTEN2 CO($J$ = 2--1), respectively.}
\label{herschel}
\end{figure}

The H{\sc ii} region RCW~36 (also known as Gum~20 or BBW~217) is one of several in the Vela molecular ridge (VMR), which is bright in the infrared. Figure \ref{herschel} shows an infrared tricolor image of RCW~36 obtained from $Herschel$ (e.g., \cite{2011A&A...533A..94H}). The red, green, and blue images represent 250-$\mu$m, 160-$\mu$m, and 70-$\mu$m observations, respectively. We clearly see a filamentary dust lane elongated from the southeast to the north of RCW~36 in the 250-$\mu$m image, while the 70-$\mu$m image corresponds to the bipolar, hourglass-shaped warm dust environment. The cluster that forms this H{\sc ii} region contains $\sim$350 number of stars including one O9V and one O9.5V. {Ages and distances of these stars are estimated to be $1.1 \pm 0.6$ Myr and  0.7 kpc, respectively \citep{2013A&A...558A.102E,1992A&A...265..577L}.} Considerable research has been devoted to the physical properties of high-mass stars, young stellar objects (YSOs), and Herbig-Haro objects located in RCW~36 using photometric and spectroscopic techniques at the optical and infrared wavelength (e.g., \cite{1994A&A...284..936V,2003A&A...399..147M,2004ApJ...614..818B,2004A&A...427L..13B,2005A&A...440..121B,2006A&A...455..561B,2012A&A...539A.156G,2011ApJ...732L...9E,2013A&A...551A...5E,2013A&A...558A.102E}). 

Numerous attempts have been made to observe the molecular clouds in the direction of RCW~36 at radio wavelengths. \citet{1977PASAu...3..147W} first observed the H$_2$CO absorption line using the Parkes 64-m radio telescope, which has an angular resolution of 4\farcm4. They found an elongated molecular cloud in the velocity range of $V_\mathrm{LSR}$ = 3--9 km s$^{-1}$, which corresponds to the dust lane distributed from the southeast to the north. \citet{1984A&A...139..181B} and \citet{1989A&AS...80..149W} found that there are two velocity components, at $V_\mathrm{LSR}$ = 3.3--5.5 km s$^{-1}$ and $V_\mathrm{LSR}$ = 7.0--9.0 km s$^{-1}$, by observing the $^{12}$CO($J$ = 1--0, 2--1) emission lines: the CO spectra obtained 0\fdg5 south of RCW~36 are possibly a blend these two components. \citet{1988A&AS...73...51M} and \citet{1991A&A...247..202M} mapped the $^{12}$CO($J$ = 1--0) line emission over the entire VMR using the Columbia University 1.2-m radio telescope, which has a relatively coarse angular resolution of 8\farcm8. They determined the CO mass of the VMR-C region (Vela C), which contains RCW~36, to be 2.0--3.7 $\times$ $10^5$ $M_\odot$, assuming a distance of 1 kpc. They also argued that the large abundance ratio of H$_2$CO / H$_2$ toward RCW~36 could be explained by the presence of shocks powered by an embedded energy source, triggering star formation in Vela C. \citet{1999PASJ...51..775Y} and \citet{2001PASJ...53.1025M} determined that the distribution of the $^{12}$CO($J$ = 1--0), $^{13}$CO($J$ = 1--0), and C$^{18}$O($J$ = 1--0) emission lines toward the VMR using the NANTEN 4-m telescope, with a finer angular resolution of $156''$. They identified 13 dense molecular clumps of C$^{18}$O within Vela C and revealed the filamentary structure of the dense molecular gas toward RCW~36. They also claimed that Vela C is the youngest star-forming region in the VMR with an age of 1 Myr or less, since there is no evolved H{\sc ii} region. Recently, \citet{2014ApJ...797L..17L} obtained $^{13}$CO($J$ = 1--0) and [C{\sc i}] $^3P_1$--$^3P_0$ maps northeast of RCW~36 using the Mopra and NANTEN2 telescopes. They found good spatial correspondence between $^{13}$CO($J$ = 1--0) and [C{\sc i}] and determined the H$_2$ column density to be $\sim$ 2 $\times$ 10$^{22}$ cm$^{-2}$ northeast of RCW~36.

In contrast, very few attempts have been made to explore the origin of the star cluster. \citet{2002aprm.conf..145H} proposed that the stellar system in RCW~36 was possibly created by a cloud-cloud collision. They argued that RCW~36 satisfies three major tests for a cloud-cloud collision model \citep{1995AJ....110.2256V}: (1) the presence of two clouds adjacent in spatial and velocity space; (2) good spatial correspondence between the infrared and $^{13}$CO peaks; and (3) a double-peaked spectrum of CO seen at the collision site. However, follow-up studies have not previously been performed.

In the present study, we use new CO datasets obtained with NANTEN2, ASTE, and Mopra to investigate whether the formation of the high-mass star cluster in RCW~36 may have been triggered by a cloud-cloud collision. The NANTEN2 CO($J$ = 1--0, 2--1) data cover a large area of RCW~36 at an angular resolution of $\sim$1\farcm5--$3'$, while the ASTE $^{12}$CO($J$ = 3--2) and Mopra $^{12}$CO($J$ = 1--0) data resolve small-scale structures with a finer angular resolution of $26''$--$45''$. Section \ref{sec:obs} describes the CO observations. Section \ref{sec:results} includes five subsections. In Subsection \ref{sec:spec}, we describe the CO spectra in the direction of RCW~36; Subsections \ref{sec:large} and \ref{sec:detail} present the large- and small-scale CO distributions; we discuss the physical properties of CO in Subsection \ref{sec:ratio}; and Subsection \ref{sec:infrared} presents a detailed comparison between the CO and infrared observations. We discuss the results and summarize our conclusions in Sections \ref{sec:discuss} and \ref{sec:conc}, respectively.

\section{Observations}\label{sec:obs}
\subsection{NANTEN2 CO $J$= 1--0 $\&$ CO $J$ = 2--1}
We performed $^{12}$CO($J$ = 1--0, 2--1) and $^{13}$CO($J$ = 2--1) observations with the NANTEN2 4-m radio telescope of Nagoya University located at Pampa La Bola ($\sim$4,865 m above sea level), in Chile.

Observations of the $^{12}$CO($J$ = 1--0) emission line at 115 GHz were conducted in May 2012. We observed an area of 2$^{\circ}$ $\times$ 1$^{\circ}$, using a Nyquist sampled on-the-fly (OTF) mapping mode. The frontend was a 4-K cooled Nb superconductor-insulator-superconductor (SIS) mixer receiver. The system temperature including the atmosphere was $\sim$190 K in the double-side band (DSB). The backend was a digital Fourier-transform spectrometer (DFS) with 16,384 channels (1 GHz band), corresponding to a velocity resolution of 0.16 km s$^{-1}$ and a velocity coverage of 2,600 km s$^{-1}$. After convolution with a 2-dimensional Gaussian kernel of $90''$, the final beam size was $\sim$180$''$ (FWHM). The absolute intensity was calibrated by observing Orion-KL [$\alpha_\mathrm{J2000}$ = $5^{\mathrm{h}}35^{\mathrm{m}}14\fs48$, $\delta_\mathrm{J2000}$ = $-5{^\circ}22\arcmin27\farcs55$] and IRAS 16293$-$2422 [$\alpha_\mathrm{J2000}$ = $16^{\mathrm{h}}32^{\mathrm{m}}23.3^{\mathrm{s}}$, $\delta_\mathrm{J2000}$ = $-24{^\circ}28\arcmin39\farcs2$] \citep{2006AJ....131.2921R}. The pointing accuracy was better than $10''$, verified by observing the edge of Sun. The typical noise fluctuation in the final data cube was $\sim$1.3 K at a velocity resolution of $\sim$0.16 km s$^{-1}$.

Observations of the $^{12}$CO($J$ = 2--1) and $^{13}$CO($J$ = 2--1) emission lines at 230 GHz were carried out simultaneously during October and November 2015. Again, we used the Nyquist sampled OTF mapping mode, and the observation area was $30'$ $\times$ $30'$ centered at ($l$, $b$) = (265\fdg17, 1\fdg47) (the solid box in Figure \ref{herschel}). The system temperature, including the atmosphere was $\sim$140--260 K in the DSB. The velocity resolution and coverage were 0.08 km s$^{-1}$ and 1,300 km s$^{-1}$, respectively. After convolution with a 2-dimensional Gaussian kernel of $45''$, the final beam size was $\sim$90$''$ (FWHM). We observed Orion-KL [$\alpha_\mathrm{J2000}$ = $5^{\mathrm{h}}35^{\mathrm{m}}14\fs48$, $\delta_\mathrm{J2000}$ = $-5{^\circ}22\arcmin27\farcs55$] to calibrate the absolute intensity by making comparisons with the CO($J$ = 2--1) dataset published by \citet{2015ApJS..216...18N}. The pointing accuracy achieved was better than $10''$. The typical noise fluctuation in the final data cube was $\sim$0.45 K for $^{12}$CO($J$ = 2--1) and $\sim$0.40 K for $^{13}$CO($J$ = 2--1) at a velocity resolution of $\sim$0.11 km s$^{-1}$.

\subsection{ASTE $^{12}$CO $J$ = 3--2}
Observations of the $^{12}$CO($J$ = 3--2) emission line at 345 GHz were carried out in November 2015 using the Atacama Submillimeter Telescope Experiment (ASTE; \cite{2004SPIE.5489..763E}), a 10-m telescope of the National Astronomical Observatory of Japan (NAOJ). We mapped an area of $16'$ $\times$ $16'$ centered at ($l$, $b$) = (265\fdg117, 1\fdg457), using the OTF mode with Nyquist sampling (dashed box in Figure \ref{herschel}). The front end was a cartridge-type, side-band-separating mixer receiver ``DASH 345.'' The typical system temperature, including the atmosphere, was 270--570 K in the single-side-band (SSB). The backend system ``MAC'' used for spectroscopy \citep{2000SPIE.4015...86S} had 1,024 channels with a total bandwidth of 128 MHz. The velocity coverage was $\sim$111 km s$^{-1}$, and the velocity resolution was $\sim$0.11 km s$^{-1}$ ch$^{-1}$. We convolved the data cube with a 2-dimensional Gaussian kernel of $11''$, giving the final beam size as $\sim$25$''$ (FWHM). The pointing accuracy was checked every hour to achieve an offset within $3''$. The absolute intensity was calibrated by observing IRC$+$10216 [$\alpha_\mathrm{B1950}$ = $9^{\mathrm{h}}45^{\mathrm{m}}14\fs8$, $\delta_\mathrm{B1950}$ = $-13{^\circ}30\arcmin40\arcsec$] \citep{1994ApJS...95..503W}, and the estimated error was less than 4 $\%$. The final data cube had a noise fluctuation of $\sim$0.29 K at a velocity resolution of 0.11 km s$^{-1}$.

\subsection{Mopra $^{12}$CO $J$ = 1--0}
To measure the intensity ratio of $^{12}$CO $J$ = 3--2 / 1--0 with a finer angular resolution of $45''$, we used the $^{12}$CO($J$ = 1--0) emission-line observations taken with the Mopra 22-m millimetre-wave telescope of the CSIRO Australia Telescope National Facility. Carried out as part of the Mopra Southern Galactic Plane CO Survey \citep{2013PASA...30...44B}, the observations were conducted between June and July 2014. The survey uses the ``fast on-the-fly'' mapping mode (see details in Section 2 of \cite{2013PASA...30...44B}), and the data used as a part of this study is an 1$^{\circ}$ $\times$ 1$^{\circ}$ area centered on ($l$, $b$) = (265\fdg5, 1\fdg0). The front end was an InP High Electron Mobility Transistor receiver. The typical system temperature was $\sim$500--900 K in the SSB. We used the UNSW Mopra Spectrometer (MOPS) system, a digital filterbank with a 4,096 channel backend with a total bandwidth of 137.5 MHz. The velocity resolution and coverage are $\sim$0.1 km s$^{-1}$ and $\sim$360 km s$^{-1}$, respectively. We convolved the data cube with a 2-dimensional Gaussian kernel of $33''$, and the final beam size was $\sim$45$''$ (FWHM). The SiO maser VY CMa was used as a pointing source for the observations, achieving an accuracy offset less than 5$''$. We also observed M17~SW [$\alpha_\mathrm{J2000}$ = $18^{\mathrm{h}}20^{\mathrm{m}}23\fs2$, $\delta_\mathrm{J2000}$ = $-16{^\circ}13\arcmin56\arcsec$], to monitor the system performance. We adopted the extended beam efficiency $\eta_\mathrm{XB}$ = 0.55 at 115 GHz to calibrate the absolute intensity \citep{2005PASA...22...62L}. The final noise fluctuation in the data cube was $\sim$1.9 K at the velocity resolution of 0.11 km s$^{-1}$.

\begin{figure}
\begin{center}
\includegraphics[width=\linewidth]{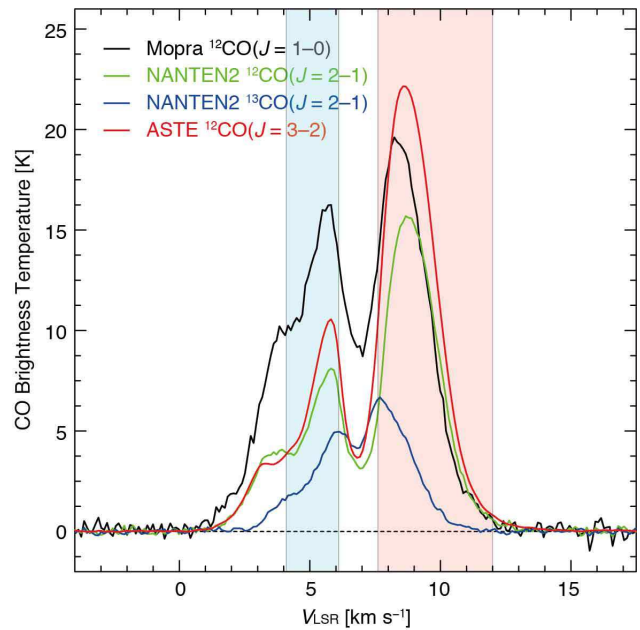}
\end{center}
\caption{CO spectra toward RCW~36. The spectra were smoothed to mach the FWHM of 90 arcsec, and averaged over a circle of 3 arcmin radius centered on the two O-type stars ($l$, $b$ = $265\fdg15$, $1\fdg45$). The black, green, blue, and red correspond to the $^{12}$CO($J$ = 1--0), $^{12}$CO($J$ = 2--1), $^{13}$CO($J$ = 2--1), and $^{12}$CO($J$ = 3--2) emission lines, respectively. The blue and red shaded areas correspond to the molecular clouds defined as blue and red clouds, respectively.}
\label{spectra}
\end{figure}%

\section{Results}\label{sec:results}
\subsection{CO spectra in the direction of RCW~36}\label{sec:spec}
Figure \ref{spectra} shows the mean CO spectra of multi-$J$-transitions in the direction of RCW~36. The $^{12}$CO emission peak at the approximate velocities $\sim$5.5 km s$^{-1}$ and $\sim$9 km s$^{-1}$, while the $^{13}$CO emission peak is at $\sim$6.1 km s$^{-1}$ and $\sim$7.6 km s$^{-1}$. The peak intensity of $^{12}$CO($J$ = 3--2) is significantly higher than that of $^{12}$CO($J$ = 1--0) and $^{12}$CO($J$ = 2--1) at 9 km s$^{-1}$, while the intensity of $^{12}$CO($J$ = 1--0) is higher than that of $^{12}$CO($J$ = 2--1) and $^{12}$CO($J$ = 3--2) at 5.5 km s$^{-1}$. We also find CO dips having a depth of 2--10 K at $\sim$7 km s$^{-1}$. It is likely that these are caused by self-absorption because the [C{\sc i}] emission line peaks at $V_\mathrm{LSR}$ $\sim$7 km s$^{-1}$ \citep{2014ApJ...797L..17L}. On the other hand, the CO distribution at $\sim$5.5 km s$^{-1}$ is completely different from that at $\sim$9 km s$^{-1}$. We hereafter refer to the component at $V_\mathrm{LSR}$ = 4.1--6.1 km s$^{-1}$ as the ``blue cloud'' and that at $V_\mathrm{LSR}$ = 7.6--12.0 km s$^{-1}$ as the ``red cloud.''

\begin{figure*}
\begin{center}
\includegraphics[width=\linewidth]{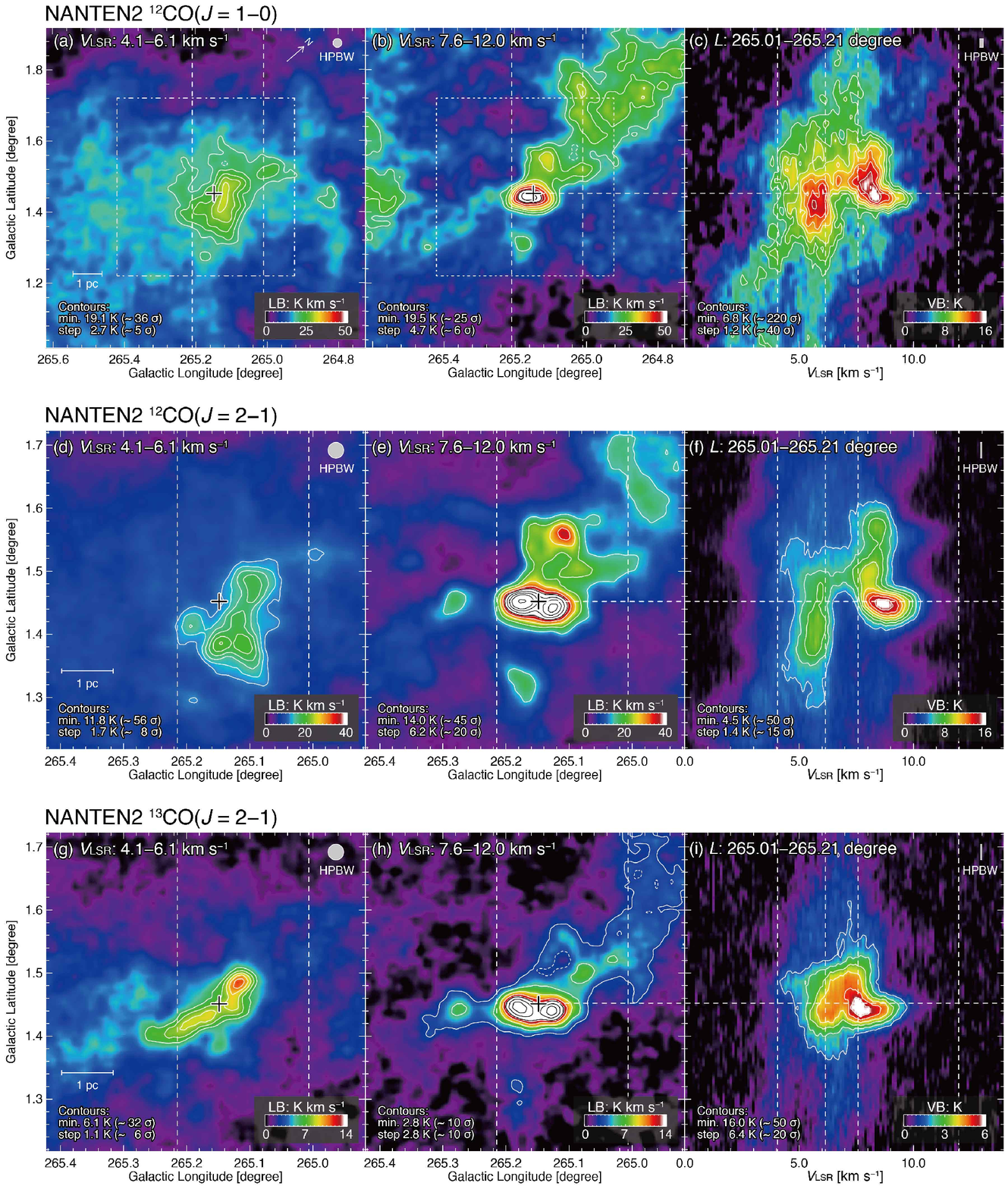}
\end{center}
\caption{Integrated intensity maps of blue and red clouds (a, b, d, e, g, h) and position-velocity diagrams (c, f, i). Upper three panels (a--c) show the $^{12}$CO($J$ = 1--0) emission line, middle three panels (d--f) show the $^{12}$CO($J$ = 2--1) emission line, and lower three panels (g--i) show the $^{13}$CO($J$ = 2--1) emission line obtained with NANTEN2. The integration ranges are shown in upper left for each panel. Vertical dashed lines indicate the integration ranges in the Galactic longitude for (a, b, d, e, g, h) and the velocity for (c, f, i). Black cross and horizontal dashed lines indicate the positions of two O-type stars \citep{2013A&A...558A.102E}.}
\label{nanten2pv}
\end{figure*}%

\begin{figure}
\begin{center}
\includegraphics[width=76mm]{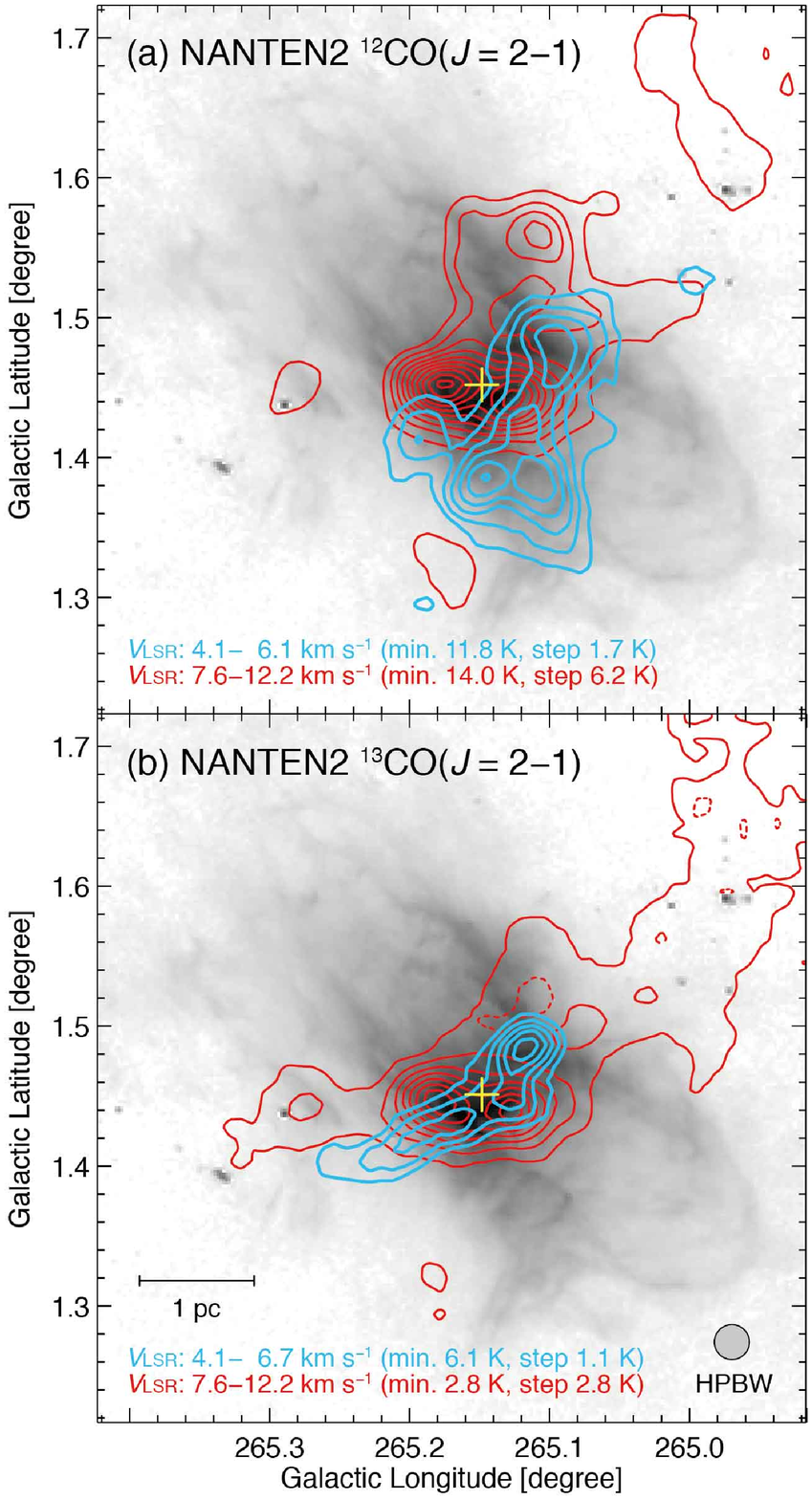}
\end{center}
\caption{Distribution of NANTEN2 (a) $^{12}$CO($J$ = 2--1) and (b) $^{13}$CO($J$ = 2--1) contours superposed on the $Herschel$ 70-$\mu$m image. Blue and red contours represent the blue and red clouds as shown in Figure \ref{nanten2pv}. The integration ranges and contour levels are the same as in Figure \ref{nanten2pv}. The position of two O-type stars red{is} also plotted in the yellow cross.}
\label{nanten2_overlay}
\end{figure}%

\subsection{Large-scale CO distribution with NANTEN2}\label{sec:large}
Figures \ref{nanten2pv}(a) and \ref{nanten2pv}(b) show the spatial distribution of $^{12}$CO($J$ = 1--0) around RCW~36. We identified two molecular clouds with different spatial distributions. The blue cloud is mainly diffuse emission extending over a region {8 pc $\times$ 4 pc} in size around RCW~36, while the red cloud has a very strong peak exactly coinciding with RCW~36 (hereafter termed the ``head''), together with a feature elongated toward the north (hereafter termed the ``tail''). The head-tail structure was also seen in the under-sampled CO data obtained using the position-switching technique with NANTEN \citep{1999PASJ...51..775Y,2001PASJ...53.1025M}. 

\begin{figure*}
\begin{center}
\includegraphics[width=\linewidth]{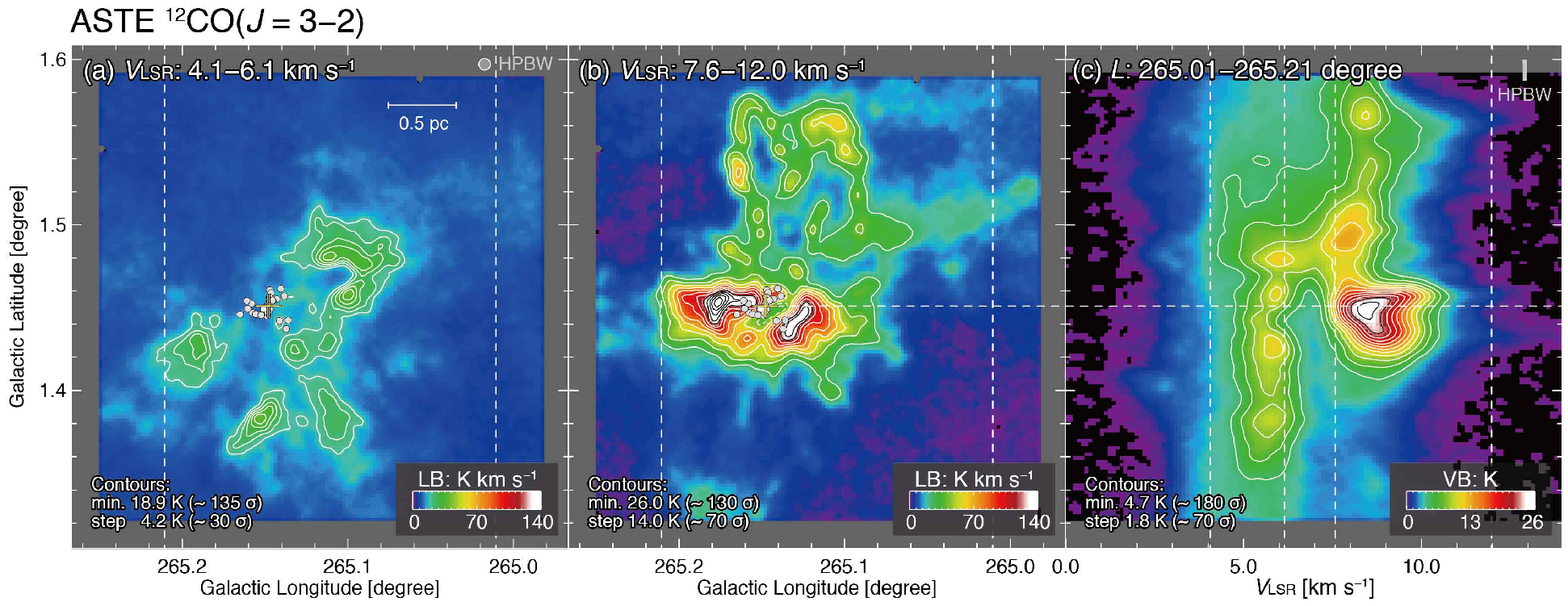}
\end{center}
\caption{(a,b) Integrated intensity maps of blue- and red clouds and (c) position-velocity diagrams of $^{12}$CO($J$ = 3--2) emission line obtained with ASTE. The integration ranges and dashed lines are the same as in Figure \ref{nanten2pv}. The yellow cross, filled circles, and filled square represent the positions of O-type stars, B-type stars, and YSOs cataloged by \citet{2013A&A...558A.102E}, respectively.}
\label{astepv}
\end{figure*}%

We obtained the masses of the blue and red clouds from 

\begin{eqnarray}
M = \mu m_{\mathrm{p}} \sum_{i} [d^2 \Omega N_i(\mathrm{H}_2)],
\label{eq1}
\end{eqnarray}

where $M$ is the mass of the molecular cloud, $\mu$ is the mean molecular weight, $m_{\mathrm{p}}$ is the mass of a proton, $d$ is the source distance, $\Omega$ is the solid angle of pixel $i$, and $N_i(\mathrm{H}_2)$ is the column density of molecular hydrogen for pixel $i$. We used $\mu$ = 2.8 to take into account the helium abundance of 20$\%$ by number relative to molecular hydrogen. We estimated the column density of molecular hydrogen using

\begin{eqnarray}
N(\mathrm{H}_2) = X \cdot W(^{12}\mathrm{CO}),
\label{eq2}
\end{eqnarray}

where $X$ is a conversion factor and $W(^{12}\mathrm{CO})$ is the integrated intensity of the $^{12}$CO($J$ = 1--0) emission line; we used the conversion factor $X$ = 1.0 $\times$ 10$^{20}$ cm$^{-2}$ (K km s$^{-1}$)$^{-1}$ \citep{2017ApJ...838..132O}. In this way we determined the molecular masses of the blue and red clouds to be $\sim$5 $\times$ 10$^3$ $M_{\odot}$ and $\sim$1 $\times$ $10^3$ $M_{\odot}$, respectively. The maximum $N$(H$_2$) of the blue cloud was found to be $\sim$3 $\times$ 10$^{21}$ cm$^{-2}$, while that of the red cloud is $\sim$7 $\times$ 10$^{21}$ cm$^{-2}$.

To evaluate influence of the self-absorption, we interpolated the absorption dip by fitting a Gaussian function for each pixel. Finally, we determined the molecular masses at $V_\mathrm{LSR}$ = 4.1--7.0 km s$^{-1}$ and 7.0--12.0 km s$^{-1}$ to be {$\sim$8 $\times$ $10^2$ $M_{\odot}$ and $\sim$2 $\times$ $10^2$ $M_{\odot}$}, respectively. The maximum $N$(H$_2$) of the blue cloud was found to be $\sim$5 $\times$ 10$^{21}$ cm$^{-2}$, while that of the red cloud is $\sim$8 $\times$ 10$^{21}$ cm$^{-2}$ (see Table \ref{table:extramath}). These values are almost the same as that of blue and red clouds. It is therefore that the self-absorption effect is almost negligible for estimation of these physical properties.

Figure \ref{nanten2pv}(c) shows the Galactic latitude--velocity diagram toward RCW~36. The blue cloud can be identified separately from the red cloud, particularly at Galactic latitudes of 1\fdg4 or smaller. Each cloud has an intensity peak near the latitude of the star cluster containing the two O-type stars. 

\begin{figure}
\begin{center}
\includegraphics[width=76mm]{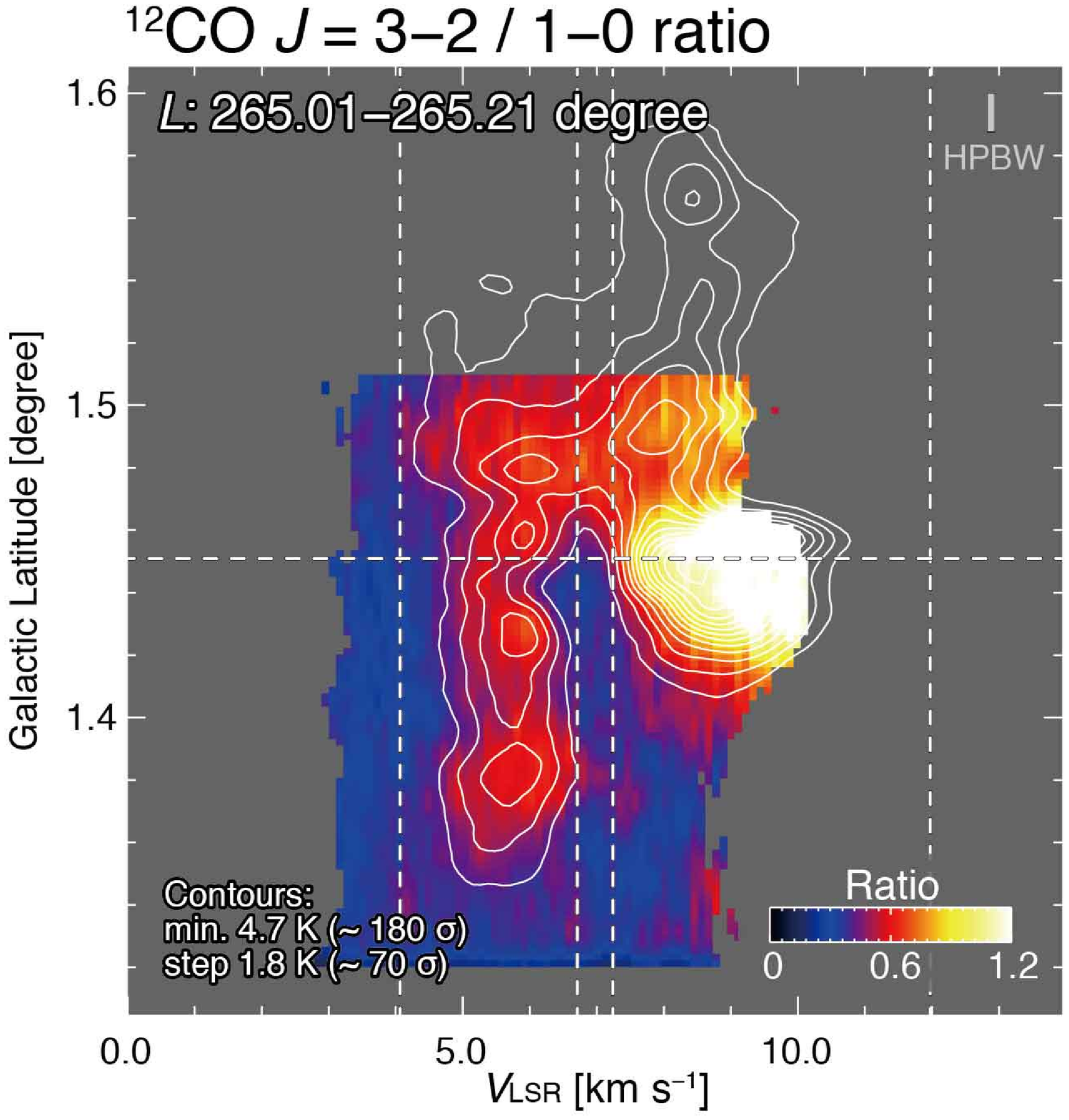}
\end{center}
\caption{Intensity ratio of $^{12}$CO $J$ = 3--2 / 1--0 using ASTE and Mopra. The integration range, superposed contours, and dashed lines are the same as Figure \ref{astepv}c.}
\label{ratio}
\end{figure}%

\begin{figure*}
\begin{center}
\includegraphics[width=\linewidth]{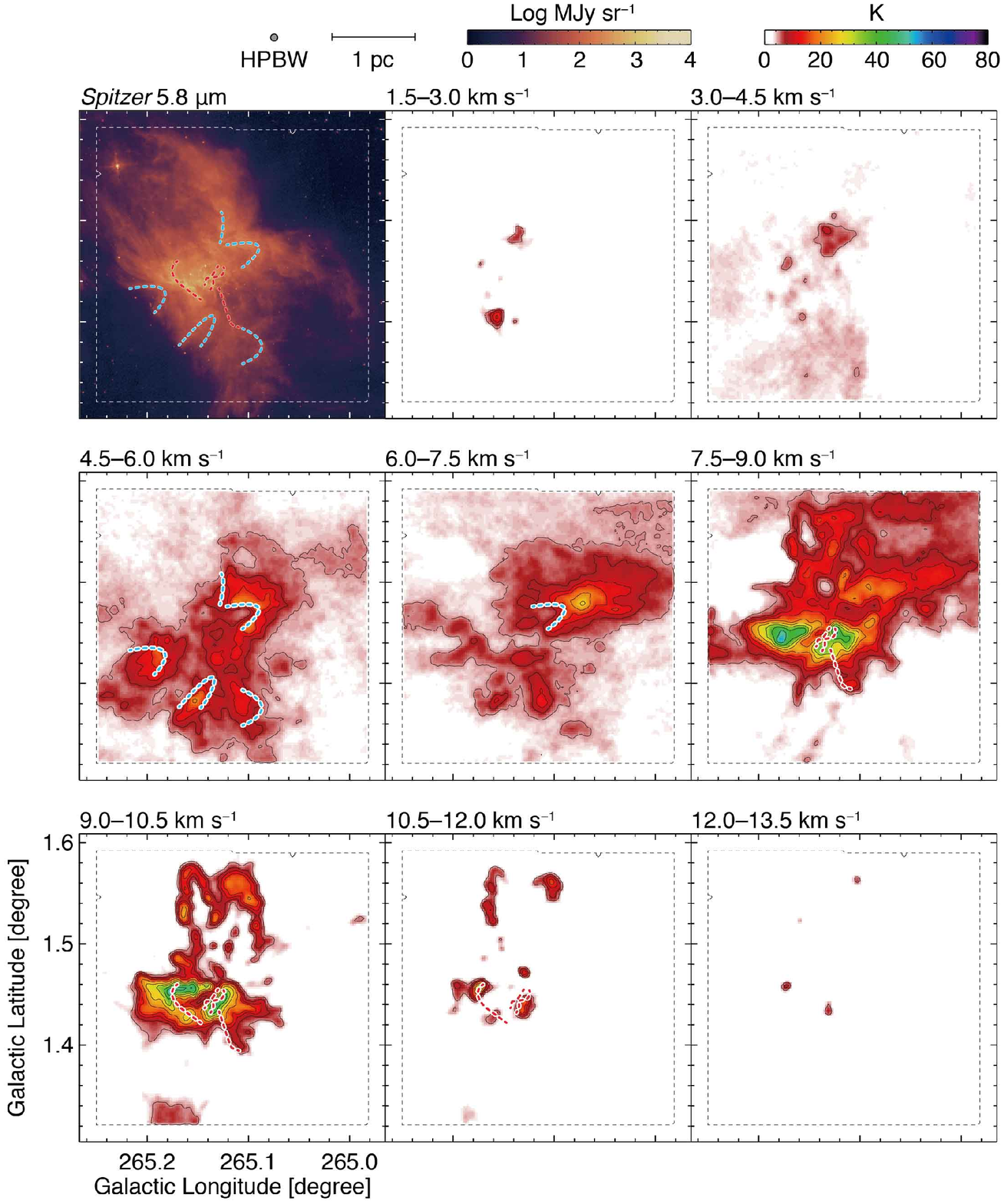}
\end{center}
\caption{Velocity channel distribution of ASTE $^{12}$CO($J$ = 3--2) and the Spitzer 5.8-$\mu$m images. Each panel shows a brightness temperature averaged every 1.5 km s$^{-1}$ in the velocity range from 1.5 km s$^{-1}$ to 13.5 km s$^{-1}$. The contour levels are 5, 8, 11, 14, 20, 26, 32, 40, and 48 K.}
\label{astechannela}
\end{figure*}%

Figures \ref{nanten2pv}(d)--\ref{nanten2pv}(i) show contour maps of $^{12}$CO($J$ = 2--1) and $^{13}$CO($J$ = 2--1) toward RCW~36. The spatial and velocity structures of $^{12}$CO($J$ = 2--1) are generally similar to those of $^{12}$CO($J$ = 1--0). Thanks to the finer angular resolution of $^{12}$CO($J$ = 2--1) ($\delta \theta$ $\sim$90$''$), we find that the head of the red cloud can be separated into two clumps, with the star cluster located in the inter-clump region. Furthermore, the cluster is surrounded by diffuse CO emission at $V_\mathrm{LSR}$ = 4.1--6.1 km s$^{-1}$. This spatial correspondence strongly suggests a physical relation between the red/blue clouds and the star cluster.

On the other hand, the distribution of $^{13}$CO($J$ = 2--1) is slightly different from that of $^{12}$CO. In Figure \ref{nanten2pv}(g), the blue cloud shows a significant feature elongated toward the south of RCW~36, while the western rim seen in $^{12}$CO($J$ = 2--1) around ($l$, $b$) = (265\fdg14, 1\fdg38) was not bright in $^{13}$CO($J$ = 2--1). In contrast, the red cloud shows a similar distribution to that of the $^{12}$CO($J$ = 2--1) emission, except for the eastern rim at ($l$, $b$) = (265\fdg15, 1\fdg55). We also find that the position-velocity diagram of $^{13}$CO in Figure \ref{nanten2pv}(i) peaks at $\sim$7 km s$^{-1}$ and within $\sim$1 pc from the star cluster in Galactic longitude, while that of $^{12}$CO in Figures \ref{nanten2pv}(c) and \ref{nanten2pv}(f) shows a dip at the same velocity due to self-absorption. This suggests that the density of the molecular cloud is higher in the central region than in the outer rim of RCW~36.

Figures \ref{nanten2_overlay}(a) and \ref{nanten2_overlay}(b) show the distribution of CO($J$ = 2--1) contours superposed on the $Herschel$ 70-$\mu$m image. It is remarkable that the local peak of the red cloud is complementary to that of the blue cloud on a pc scale. We also find that the $^{13}$CO clouds are elongated from south to north, while the bipolar infrared filaments at 70-$\mu$m are distributed perpendicular to the $^{13}$CO clouds.

\subsection{Detailed CO distribution with ASTE}\label{sec:detail}
Figures \ref{astepv}(a) and \ref{astepv}(b) show the spatial distributions of $^{12}$CO($J$ = 3--2) for the blue and red clouds obtained using ASTE, with an angular resolution of $25''$. The blue cloud can be categorized as four molecular clumps with sizes of {$\sim$0.5--1 pc}, while the region located at the star cluster has no dense molecular clumps associated with it. In contrast, the star cluster is embedded within two dense molecular clumps in the red cloud, and YSOs cataloged by \citet{2013A&A...558A.102E} are also located along the edge of this molecular cloud (Figure \ref{astepv}b). We also newly identify two molecular features toward the northeast of RCW~36 that have bow-shock-like structures, with sizes of {1.5 pc $\times$ 0.5 pc} for the left one and {1.5 pc $\times$ 0.7 pc} for the right one in Galactic longitude. The typical thickness of the bow-shock-like structures is estimated to be {$\sim$0.1 pc}. The CO structures also show a velocity gradient from $\sim$7.6 km s$^{-1}$ to $\sim$9.6 km s$^{-1}$. To understand the origins of these structures is not the main topic of the present study, and we do not discuss it further in this paper. We expect that ALMA observations, which have a finer angular resolution of $1''$ or less, together with deep-infrared observations will enable us us to understand the origin of the bow-shock-like structures.

Figure \ref{astepv}c shows the position-velocity diagram of $^{12}$CO($J$ = 3--2). The integration range is the same as in Figures \ref{nanten2pv}(c), \ref{nanten2pv}(f), and \ref{nanten2pv}(i). The red cloud has the highest intensity toward the star cluster, while the blue cloud shows a dip around ($V_\mathrm{LSR}$, $b$) $\sim$ (6 km s$^{-1}$, 1\fdg4).

\subsection{Intensity ratio of $^{12}$CO $J$ = 3--2 / 1--0}\label{sec:ratio}
To understand the physical conditions in the molecular clouds, we consider the intensity ratio between different $J$-transitions of CO (e.g., \cite{2010ApJ...709..975O, 2015ApJ...806....7T, 2016ApJ...820...26F}). Figure \ref{ratio} shows the intensity ratio of $^{12}$CO $J$ = 3--2 / 1--0 in the position-velocity diagram. According to \citet{2012ApJS..201...14O}, the typical intensity ratio for a dark cloud is 0.4 or lower. In contrast, we clearly see that the intensity ratio is $\sim$0.6--1.2 in the direction of RCW~36. Since the intensity ratio of the red cloud and the component at $V_\mathrm{LSR}$ $\sim$7 km s$^{-1}$ are strongly affected by the self-absorption of the $^{12}$CO($J$ = 1--0) emission, these intensity ratios are unreliable. The blue cloud also shows a relatively high intensity ratio of $\sim$0.6, although the southern part of this cloud, at Galactic longitude $<$ 1\fdg4, has a significantly lower intensity ratio than the red cloud.

\subsection{Detailed comparison with infrared observations}\label{sec:infrared}
Figure \ref{astechannela} shows the velocity-channel map of $^{12}$CO($J$ = 3--2) and the $Spitzer$ 5.8-$\mu$m image. We identified eight infrared filaments, which show good spatial correspondence with the CO cloud. The blue and red dashed lines represent features of the blue cloud ($V_\mathrm{LSR}$ =  4.1--6.1 km s$^{-1}$) and of red cloud ($V_\mathrm{LSR}$ = 7.6--12.0 km s$^{-1}$), respectively. The infrared features corresponding to the blue cloud are distributed in the outer regions of the infrared bipolar nebula. In contrast, the brightest infrared features located in the central region of RCW~36 show good spatial correspondence with the edge of red cloud. We also find high-velocity CO clumps around the star cluster in the velocity ranges of $V_\mathrm{LSR}$ = 1.5--3.0 km s$^{-1}$ and $V_\mathrm{LSR}$ = 12.5--13.5 km s$^{-1}$.

\section{Discussion} \label{sec:discuss}
\subsection{Two molecular clouds associated with RCW~36}
The present CO observations have revealed two molecular clouds, which we have termed the blue and red clouds, with a velocity separation of $\sim$3.5 km s$^{-1}$ in the direction of RCW~36. We argue that both clouds are physically associated with the star cluster in RCW~36. On the large scale of {$\sim$1 pc}, both CO clouds show good spatial correspondence with the position of the star cluster (Figure \ref{nanten2pv}). We also found that the red cloud coincides with the YSOs and O-type stars {at $\sim$0.1 pc} resolution (Figure \ref{astepv}), which is characteristic of star-forming regions (e.g., \cite{2008hsf1.book..483M,2008hsf2.book..599O,2008hsf2.book..625C}). The blue cloud also shows a dip-like structure in Figure \ref{astepv}c, suggestive of ionization by the UV radiation from the star cluster. In addition, the intensity ratio of CO $J$ = 3--2 / 1--0 toward both the blue and red clouds shows a significantly higher value $\sim$0.6--1.2 than that typical of a dark cloud (e.g., \cite{2012ApJS..201...14O}), indicating that the gas temperature has been increased due to heating by the O-type stars. Moreover, the eight infrared filaments coincide with the edges of the blue cloud and the densest region of the red cloud. We therefore conclude that both the blue and red clouds are physically associated with the star cluster and RCW~36.

\subsection{High-mass star cluster formation triggered by cloud-cloud collision}
\citet{2017arXiv170104669F} argued that the observational signatures of O-type star formation by cloud-cloud collisions are categorized by three elements: (1) the two colliding molecular clouds show a supersonic velocity separation, (2) a bridging feature connects the two clouds in velocity space, and (3) the clouds display complementary spatial distributions. To demonstrate that high-mass star formation in RCW~36 is triggered by such a cloud-cloud collision, we compare these signatures with the observational results for RCW~36. In this section, we show that the cloud-cloud collision scenario is likely to be the mechanism that created the high-mass star cluster in RCW~36.

In RCW~36, the observed velocity separation of the two clouds is $\sim$3.5 km s$^{-1}$. Table \ref{table:extramath} summarizes the physical properties of the molecular clouds associated with high-mass stars formed by cloud-cloud collisions. This small velocity separation is similar to the case of the Orion Nebula Cluster (ONC), for which \citet{2017arXiv170104669F} found that M42 and M43 were also created by cloud-cloud collisions with a velocity separation of 7 km s$^{-1}$. We emphasize that these velocity separations cannot be created by the stellar winds and UV-radiation from the high-mass stars.

\begin{table*}[h]
\tbl{Physical properties of molecular clouds toward H{\sc ii} regions formed by the cloud-cloud collision}{%
\begin{tabular}{lccccccccc} 
\hline 
\hline\noalign{\vskip3pt} 
\multicolumn{1}{c}{Name} & Molecular mass & $N$(H$_2$) & Relative & Complementary & Bridging & Age  & Number of & Cluster & References \\ [-3pt]
& & & velocity & distribution & feature & &  O-type stars & type &  \\ 
 & [$M_\odot$] & [cm$^{-2}$] & [km s$^{-1}$] &  &  & [Myr] &  & &  \\  
\multicolumn{1}{c}{(1)} & (2) & (3) & (4) & (5) & (6) & (7) & (8) & (9) & (10) \\  [2pt] 
\hline\noalign{\vskip3pt} 
\multirow{2}{*}{RCW~38}  & $2 \times 10^4$  &  $1 \times 10^{23}$ & \multirow{2}{*}{12} & \multirow{2}{*}{no} & \multirow{2}{*}{yes} & \multirow{2}{*}{$\sim 0.1$} & \multirow{2}{*}{$\sim 20$} & \multirow{2}{*}{SSC} & \multirow{2}{*}{[1]}    \\[-3pt] 
  & $3 \times 10^3$  & $1 \times 10^{22}$ &  &  &  &  &  &   \\  [2pt] 
\multirow{2}{*}{NGC~3603}  & $7 \times 10^4$  &  $1 \times 10^{23}$ & \multirow{2}{*}{15} & \multirow{2}{*}{no} & \multirow{2}{*}{yes} & \multirow{2}{*}{$\sim 2.0$} & \multirow{2}{*}{$\sim 30$} & \multirow{2}{*}{SSC} & \multirow{2}{*}{[2]}  \\[-3pt] 
  & $1 \times 10^4$  & $1 \times 10^{22}$ &  &  &  &  &  &   \\  [2pt] 
\multirow{2}{*}{Westerlund~2}  & $9 \times 10^4$  &  $2 \times 10^{23}$ & \multirow{2}{*}{16} & \multirow{2}{*}{yes} & \multirow{2}{*}{yes} & \multirow{2}{*}{$\sim 2.0$} & \multirow{2}{*}{14} & \multirow{2}{*}{SSC} & \multirow{2}{*}{[3, 4]}  \\[-3pt] 
  & $8 \times 10^4$  & $2 \times 10^{22}$ &  &  &  &  &  &   \\  [2pt] 
\multirow{2}{*}{[DBS2003]~179}  & $2 \times 10^5$  &  $8 \times 10^{22}$ & \multirow{2}{*}{20} & \multirow{2}{*}{yes} & \multirow{2}{*}{yes} & \multirow{2}{*}{$\sim 5.0$} & \multirow{2}{*}{$> 10$} & \multirow{2}{*}{SSC} & \multirow{2}{*}{[5]}  \\[-3pt] 
  & $2 \times 10^5$  & $5 \times 10^{22}$ &  &  &  &  &  &   \\ [2pt] 
\hline\noalign{\vskip3pt} 
\multirow{2}{*}{M42}  & $2 \times 10^4$  &  $2 \times 10^{23}$ & \multirow{2}{*}{7$^\dagger$} & \multirow{2}{*}{yes} & \multirow{2}{*}{no} & \multirow{2}{*}{$\sim 0.1$} & \multirow{2}{*}{$\sim 10$} &\multirow{2}{*}{SC}& \multirow{2}{*}{[6]}  \\[-3pt] 
  & $3 \times 10^3$  & $2 \times 10^{22}$ &  &  &  &  &  &   \\  [2pt] 
\multirow{2}{*}{M43}  & $3 \times 10^2$  &  $6 \times 10^{22}$ & \multirow{2}{*}{7$^\dagger$} & \multirow{2}{*}{yes} & \multirow{2}{*}{no} & \multirow{2}{*}{$\sim 0.1$} & \multirow{2}{*}{$\sim 1$} &\multirow{2}{*}{SC}& \multirow{2}{*}{[6]}  \\[-3pt] 
  & $2 \times 10^2$  & $2 \times 10^{22}$ &  &  &  &  &  &   \\  [2pt] 
\multirow{2}{*}{RCW~36}  & {$8\times 10^2$}  &  $3 \times 10^{21}$ & \multirow{2}{*}{5$^\dagger$} & \multirow{2}{*}{yes} & \multirow{2}{*}{no} & \multirow{2}{*}{{$\sim 0.1$}} & \multirow{2}{*}{2} &\multirow{2}{*}{SC}& \multirow{2}{*}{This {study}}  \\[-3pt] 
  & {$2 \times 10^2$}  & $7 \times 10^{21}$ &  &  &  &  &  &   \\ [2pt] 
\hline\noalign{\vskip3pt} 
\multirow{2}{*}{M20}  & $1 \times 10^3$  &  $1 \times 10^{22}$ & \multirow{2}{*}{7.5} & \multirow{2}{*}{yes} & \multirow{2}{*}{yes} & \multirow{2}{*}{$\sim 0.3$} & \multirow{2}{*}{1} &\multirow{2}{*}{-----}& \multirow{2}{*}{[7]}  \\[-3pt] 
  & $1 \times 10^3$  & $1 \times 10^{22}$ &  &  &  &  &  &   \\  [2pt] 
\multirow{2}{*}{RCW~120}  & $5 \times 10^4$  &  $3 \times 10^{22}$ & \multirow{2}{*}{20} & \multirow{2}{*}{yes} & \multirow{2}{*}{yes} & $\sim 0.2$ & \multirow{2}{*}{1} &\multirow{2}{*}{-----}& [8]  \\[-3pt] 
  & $4 \times 10^3$  & $8 \times 10^{21}$ &  &  &  & $< 5.0$ & & & [9]  \\  [2pt] 
\multirow{2}{*}{N159W-South}  & $9 \times 10^3$  &  $1 \times 10^{23}$ & \multirow{2}{*}{8$^\dagger$} & \multirow{2}{*}{no} & \multirow{2}{*}{no} & \multirow{2}{*}{$\sim 0.06$} & \multirow{2}{*}{1} &\multirow{2}{*}{-----}& \multirow{2}{*}{[10]}  \\[-3pt] 
  & $6 \times 10^3$  & $1 \times 10^{23}$ &  &  &  &  &  &   \\  [2pt] 
\multirow{3}{*}{N159E-Papillon}  & $5 \times 10^3$  &  $4 \times 10^{22}$ & \multirow{3}{*}{9$^\dagger$} & \multirow{3}{*}{no} & \multirow{3}{*}{no} & \multirow{3}{*}{$\sim 0.2$} & \multirow{3}{*}{1} &\multirow{3}{*}{-----}& \multirow{3}{*}{[11]}  \\[-3pt] 
  & $7 \times 10^3$  & $4 \times 10^{22}$ &  &  &  &  &  &   \\[-3pt] 
   & $8 \times 10^3$  & $6 \times 10^{22}$ &  &  &  &  &  &   \\  [2pt]   
\hline
\hline\noalign{\vskip3pt} 
\end{tabular}}
\label{table:extramath}
\begin{tabnote}
\hangindent6pt\noindent
Note. --- Col. (1): H{\sc ii} region name. Cols. (2--3) Physical properties of collided two or three clouds. Col. (2) Molecular mass. Col. (3) Molecular column density $N$(H$_2$). Cols. (4--6) Relation among the two or three clouds. Col. (4) Relative velocity separation. Col. (5) Complementary spatial distribution. Col. (6) Bridging feature in the velocity space. Col. (7) Age of the cluster or O-type star. Col. (8) Number of O-type stars. Col. (9) Cluster type. SSC and SC indicate Super Star Cluster and Star Cluster. Col. (10) References. [1] \cite{2016ApJ...820...26F}, [2] \cite{2014ApJ...780...36F}, [3] \cite{2009ApJ...696L.115F}, [4] \cite{2010ApJ...709..975O}, [5] \cite{kuwahara_prep}, [6] \cite{2017arXiv170104669F}, [7] \cite{2011ApJ...738...46T}, [8] \cite{2010A&A...510A..32M}, [9] \cite{2015ApJ...806....7T}, [10] \cite{2015ApJ...807L...4F}, [11] \cite{2017ApJ...835..108S}.\\
$^\dagger$Corrected for the projection effect.
\end{tabnote}
\end{table*}

We did not observe a bridging feature in RCW~36, except for the self-absorbed velocity at $V_\mathrm{LSR}$ $\sim$7 km s$^{-1}$, but this is not inconsistent with the cloud-cloud collision scenario. Bridging features also are not seen in N159W-South, N159E-Papillon, or the ONC, each of which has a small velocity difference of a few km s$^{-1}$, along the line-of-sight \citep{2015ApJ...807L...4F, 2017arXiv170104669F, 2017ApJ...835..108S}. Since the two clouds associated with RCW~36 have a small velocity separation of only $\sim$3.5 km s$^{-1}$, a bridging feature is not expected for this source.

\begin{figure*}
\begin{center}
\includegraphics[width=\linewidth]{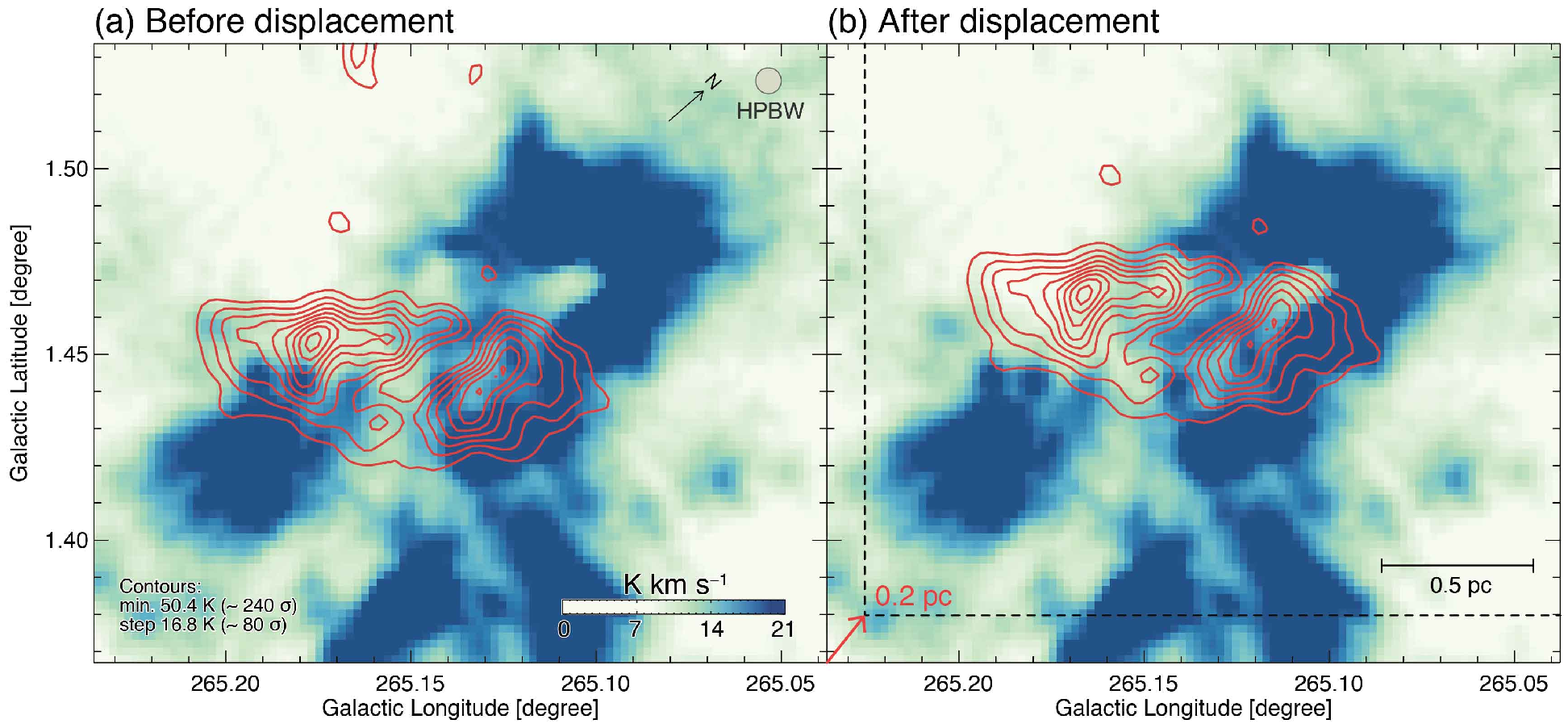}
\end{center}
\caption{Complementary $^{12}$CO($J$ = 3--2) distribution of the blue and red clouds. The image and contours indicate the blue and red clouds, respectively. The integration range is the same as in Figure \ref{astepv}. (a) and (b) correspond to the maps of before and after displacement of the red cloud contours, respectively.}
\label{displacement}
\end{figure*}%

We also note that the small number of O-type stars in RCW~36 (O9V $\&$ O9.5V; \cite{2013A&A...558A.102E}) is consistent with the low column density of their natal molecular clouds, $\sim$1 $\times$ 10$^{22}$ cm$^{-2}$. According to \citet{2017arXiv170104669F}, the formation of a super star cluster (SSC) containing 10--20 O-type stars requires a collision between clouds with high densities, $\sim$10$^{23}$ cm$^{-2}$ (e.g., RCW~38, NGC~3603, Westerlund 2), whereas the formation of just a single O-type star happens in a collision between clouds with lower column densities, around 10$^{22}$ cm$^{-2}$ (e.g., M43, M20, RCW~120). If RCW~36 did originate in a cloud-cloud collision, then it is intermediate between the SSC-forming and single-O-star-forming H{\sc ii} regions.

Next, we argue that the complementary spatial distribution of the two clouds is also seen in RCW~36. The colliding clouds generally have different sizes, as indicated in simulations by \citet{1992PASJ...44..203H}, \citet{2010MNRAS.405.1431A}, and \citet{2014ApJ...792...63T}. According to these numerical simulations, the colliding smaller cloud creates a cavity-like structure in the larger cloud. This produces complementary spatial distributions of the two molecular clouds, with different radial velocities. Figure 4 shows the complementary spatial distribution of the two molecular clouds in RCW~36, which have different velocities $V_\mathrm{LSR}$ = 4.1--6.1 km s$^{-1}$ and $V_\mathrm{LSR}$ = 7.6--12.0 km s$^{-1}$, suggestive of the characteristic structure expected for a cloud-cloud collision. Since RCW~36 thus satisfies the observational signatures of a cloud-cloud collision, we conclude that RCW~36 and the high-mass stars it contains were likely created by a cloud-cloud collision.

\subsection{Collisional time scale}
Since the angle $\theta$ of two colliding clouds is generally not at 0$^{\circ}$ or 90$^{\circ}$ relative to the line-of-sight, we can also observe the displacement between the complementary distributions of the small cloud and the cavity in the large cloud. \citet{2017arXiv170104669F} presented displacements for the cases $\theta$ = 0$^{\circ}$, 45$^{\circ}$, and 90$^{\circ}$ using the numerical results of \citet{2014ApJ...792...63T}. They found a displacement only for the case of $\theta$ = 45$^{\circ}$. Although we could not determine an accurate value of $\theta$ from our CO datasets, $\theta$ = 45$^{\circ}$ is one of the convenient solutions as a first approximation of the angle of two colliding clouds with a displacement. They also studied observationally not only the displacement of the colliding clouds in the directions of M42 and M43, but also they obtained the collision time scale by dividing the distance of the displacement by the velocity separation between the two clouds assuming $\theta$ = 45$^{\circ}$ (see also Section 4.2 of \cite{2017arXiv170104669F}). 

\begin{table}[h]
\tbl{Fitting parameters of complementary displacement}{%
\begin{tabular}{lc} 
\hline 
\hline\noalign{\vskip3pt} 
Small cloud (K km s$^{-1}$) ............. & {$> 51.4$}\\
Cavity cloud (K km s$^{-1}$) ............ & $< 19.0$\\
Integration step (pc) ................... & {0.03}\\
Position angle (degree) ............... & 3\phantom{0}[$-2$, 8]\\
$H$($\Delta$) .......................................... & {0.615\phantom{0}}[0.601, {0.613]}\\
Displacement (pc) ...................... & {0.2}\\
\hline\noalign{\vskip3pt} 
\end{tabular}}
\label{table:displacement}
\begin{tabnote}
\hangindent6pt\noindent
Note. --- Integration step is the minimum interval of the displacement. Position angle indicates the direction of the displacement. Overlap integral $H$($\Delta$) is defined as an overlap area between the small and cavity clouds, which is normalized by the area of the small cloud. 
\end{tabnote}
\end{table}

In RCW~36, we also found the displacement between the two clouds, following the same method presented by \citet{2017arXiv170104669F}. Figure \ref{displacement} shows the red cloud contours for $^{12}$CO($J$ = 3--2) superposed on the map of the blue cloud. We estimate the overlap integral $H$($\Delta$), which defined as an overlapping area between the red cloud enclosed by the lowest contour ($>$ {51.4} K km s$^{-1}$) and the hole of the blue cloud ($<$ 19 K km s$^{-1}$). We calculated $H$($\Delta$) as a function of displacement of the red cloud with {0.03} pc steps to the direction of position angle $\sim$3$^{\circ}$. We confirmed that the position angle $\sim$3$^{\circ}$ gives the maximum value of $H$($\Delta$) by changing the angle in a range of $-2^\circ$--8$^\circ$ (more details are shown in Appendix of \cite{2017arXiv170104669F}). The fitting parameters are summarized in Table \ref{table:displacement}. Finally, we obtained the projected displacement to be {$\sim$0.2 pc} toward the north, which shows the maximum value of $H$($\Delta$). {The collision timescale of RCW~36 is then given by $\sim$0.2 pc / 3.5 km s$^{-1} (\tan \theta)^{-1}$. If we adopt the $\theta$ = 15$^{\circ}$, 45$^{\circ}$, and 75$^{\circ}$, the collision timescales are to be 0.2, 0.06, and 0.02 Myr, respectively. The dynamical age of collisions $\sim$10$^5$ yr is shorter than the cluster age $\sim$1 Myr presented by \citet{2013A&A...558A.102E}.}

\begin{figure}
\begin{center}
\includegraphics[width=\linewidth]{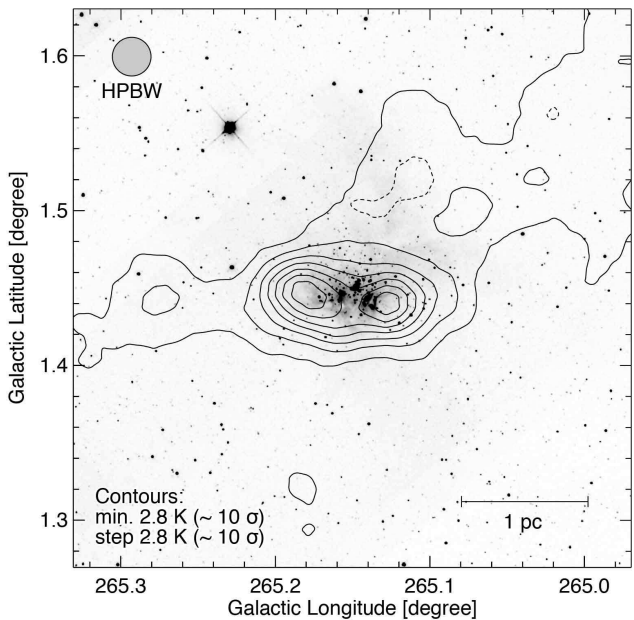}
\end{center}
\caption{{Distribution of $^{13}$CO($J$ = 2--1) contours superposed on the Two Micron All Sky Survey (2MASS) $K_\mathrm{s}$ image. Contours represent the red clouds as shown in Figure \ref{nanten2pv}h. The integration ranges and contour levels are the same as in Figure \ref{nanten2pv}h.}}
\label{lowmassstars}
\end{figure}

{We here present a possible scenario that O-type star formation is not coeval with the formation of the other low-mass members. According to \citet{2016ApJ...820...26F} and \citet{2017arXiv170104669F}, young stellar clusters in RCW~38 and M42 show similar spatial distributions of the dense molecular gas, having molecular column density of $\sim$10$^{22}$ cm$^{-2}$, and the low-mass member stars. These suggested that the low-mass stars were formed before the cloud-cloud collisions. In RCW~36, Figure \ref{lowmassstars} shows a comparison of distribution between the $K_\mathrm{s}$ band stellar objects with the $^{13}$CO($J$ =2--1) distribution and the two show a good spatial correspondence. It is probable that the cluster age by \citet{2013A&A...558A.102E} is dominated by the low-mass member stars which were not formed under the triggering by cloud-cloud collision, and that the O-type stars in the center of the cluster is explained by the collisional triggering independently from the low-mass star formation. Although the value of $\theta$ was an issue, in this scenario it is possible that the age of O-type star is as young as $\sim$10$^5$ yrs as in cases of RCW~38 and M42.}

{Finally, we discuss that $\theta$$\sim$45$^{\circ}$ provides a natural geometry where the collision path length is similar to the projected displacement between the two colliding clouds. If we adopt $\theta$ close to 0$^{\circ}$, we can obtain the  dynamical age of collisions $\sim$1 Myr. However, we have an unreasonable ad-hoc geometry where the collision path length is more than ten times longer than the displacement, the probability for which should be very small. The previous studies in RCW~38 and M42 suggest that O-type star formation is not necessarily coeval with the low-mass stars, and that, once O-type stars are formed, the ionization disperses the cloud, leading to termination of star formation at least near the O-type stars. The existence of dense molecular clouds in RCW~36 provides alternative support that the O-type stars are relatively young similar to the cases of RCW~38 and M42.}

\section{Conclusions} \label{sec:conc}
We have made new CO observations of RCW~36 with NANTEN2, Mopra, and ASTE using $^{12}$CO($J$ = 1--0, 2--1, 3--2) and $^{13}$CO($J$ = 2--1) line emissions to investigate whether the formation of the high-mass star cluster in RCW 36 may have been triggered by a cloud-cloud collision. We have shown the cloud-cloud collision model offers a reasonable scenario to explain the observed CO clouds and the star cluster properties in RCW~36. We summarize the main results of the present study as follows;

\begin{enumerate}
\item We have determined the distribution of two molecular clouds associated with the high-mass star cluster RCW~36 by using the NANTEN2, Mopra, and ATSE CO datasets. The blue cloud (at $V_\mathrm{LSR}$ = 4.1--6.1 km s$^{-1}$) consists of diffuse emission surrounding RCW~36, while the red cloud (at $V_\mathrm{LSR}$ = 7.6--12.0 km s$^{-1}$) has two strong CO peaks with a tail-like structure elongated toward the north. The masses of the red and blue clouds were estimated to be {$\sim$8 $\times$ 10$^2$ $M_\mathrm{\odot}$ and $\sim$ 2 $\times$ 10$^2$ $M_{\odot}$}, respectively.
\item It is likely that the blue and red clouds are physically associated with the high-mass star cluster in RCW~36. The clouds exhibit three characteristic features associated with the cloud-cloud collisions: (a) Both clouds show good spatial correspondence with the position of the star cluster at the {$\sim$ 1 pc scale}, while the red cloud coincides with the distributions of O-type stars and YSOs at the {$\sim$ 0.1 pc} resolution. (b) The high intensity ratio of CO $J$ = 3--2 / 1--0 $\sim$ 0.6--1.2, indicates that the gas temperature is increased due to heating by the O-type stars. (c) There is good spatial correspondence between the infrared filaments and the two molecular clouds. 
\item We argue that RCW~36 and its star cluster were formed by a collision between the red and blue clouds. The complementary spatial distributions and velocity separations of the two clouds are in good agreement with the expected observational signatures of O-star formation triggered by cloud-cloud collisions (e.g., \cite{2017arXiv170104669F}). The low column density of $\sim$ 1 $\times$ 10$^{22}$ cm$^{-2}$ is also consistent with the small number of O-type stars formed in RCW~36 (see \cite{2017arXiv170104669F} and references therein). We also found the displacement of the complementary spatial distributions of the two clouds, which we estimate to be {0.2 pc as a projected distance}. We finally proposed a scenario for the cloud-cloud collision in RCW~36, in which the smaller red cloud collided with the larger diffuse blue cloud at a relative velocity of $\sim$5 km s$^{-1}$. {We found the collision timescale to be $\sim$10$^5$ yr. It can be interpreted that the cluster age by \citet{2013A&A...558A.102E} is dominated by the low-mass members which were formed before the collision, while the O-type stars are explained by the collisional triggering independently from the low-mass star formation.}. \end{enumerate}

\begin{ack}
NANTEN2 is an international collaboration of 11 universities: Nagoya University, Osaka Prefecture University, University of Bonn, University of Cologne, Seoul National University, University of Chile, University of New South Wales, Macquarie University, University of Sydney, University of Adelaide, and University of ETH Zurich. The Mopra radio telescope is part of the Australia Telescope National Facility. The University of New South Wales, the University of Adelaide, and the National Astronomical Observatory of Japan (NAOJ) Chile Observatory supported operations. The ASTE telescope is operated by NAOJ. This {study} was financially supported by Grants-in-Aid for Scientific Research (KAKENHI) of the Japanese society for the Promotion of Science (JSPS, grant No. 15H05694). This {study} also was supported by ``Building of Consortia for the Development of Human Resources in Science and Technology'' of Ministry of Education, Culture, Sports, Science and Technology (MEXT, grant No. 01-M1-0305). This {study} was initiated by preliminary studies done by high school students, Saori Morita, Fumika Demachi, and Shizuka Hoshi who visited Nagoya University as part of Super Science Highschool (SSH). We acknowledge to Tetsuta Inaba and Ryuji Okamoto for contributions on the observations of $^{12}$CO($J$ = 3--2) data. We are also grateful to Ryosuke Asano and the anonymous referee for useful comments which helped the authors to improve the paper.
\end{ack}


\begin{thebibliography}{99}
\bibitem[Anathpindika(2010)]{2010MNRAS.405.1431A} Anathpindika, S.~V.\ 2010, \mnras, 405, 1431 
\bibitem[Baba et al.(2004)]{2004ApJ...614..818B} Baba, D., Nagata, T., Nagayama, T., et al.\ 2004, \apj, 614, 818 
\bibitem[Bik \& Thi(2004)]{2004A&A...427L..13B} Bik, A., \& Thi, W.~F.\ 2004, \aap, 427, L13 
\bibitem[Bik et al.(2005)]{2005A&A...440..121B} Bik, A., Kaper, L., Hanson, M.~M., \& Smits, M.\ 2005, \aap, 440, 121 
\bibitem[Bik et al.(2006)]{2006A&A...455..561B} Bik, A., Kaper, L., \& Waters, L.~B.~F.~M.\ 2006, \aap, 455, 561 
\bibitem[Brand et al.(1984)]{1984A&A...139..181B} Brand, J., van der Bij, M.~D.~P., de Vries, C.~P., et al.\ 1984, \aap, 139, 181
\bibitem[Burton et al.(2013)]{2013PASA...30...44B} Burton, M.~G., Braiding, C., Glueck, C., et al.\ 2013, PASA, 30, e044 
\bibitem[Chini \& Hoffmeister(2008)]{2008hsf2.book..625C} Chini, R., \& Hoffmeister, V.\ 2008, Handbook of Star Forming Regions, Volume II, 5, 625 
\bibitem[Ellerbroek et al.(2011)]{2011ApJ...732L...9E} Ellerbroek, L.~E., Kaper, L., Bik, A., et al.\ 2011, \apjl, 732, L9 
\bibitem[Ellerbroek et al.(2013a)]{2013A&A...551A...5E} Ellerbroek, L.~E., Podio, L., Kaper, L., et al.\ 2013a, \aap, 551, A5 
\bibitem[Ellerbroek et al.(2013b)]{2013A&A...558A.102E} Ellerbroek, L.~E., Bik, A., Kaper, L., et al.\ 2013b, \aap, 558, A102 
\bibitem[Ezawa et al.(2004)]{2004SPIE.5489..763E} Ezawa, H., Kawabe, R., Kohno, K., \& Yamamoto, S.\ 2004, \procspie, 5489, 763 
\bibitem[Fukui et al.(2014)]{2014ApJ...780...36F} Fukui, Y., Ohama, A., Hanaoka, N., et al.\ 2014, \apj, 780, 36 
\bibitem[Fukui et al.(2015)]{2015ApJ...807L...4F} Fukui, Y., Harada, R., Tokuda, K., et al.\ 2015, \apjl, 807, L4 
\bibitem[Fukui et al.(2016)]{2016ApJ...820...26F} Fukui, Y., Torii, K., Ohama, A., et al.\ 2016, \apj, 820, 26 
\bibitem[Fukui et al.(2017)]{2017arXiv170104669F} Fukui, Y., Torii, K., Hattori, Y., et al.\ 2017, arXiv:1701.04669 
\bibitem[Furukawa et al.(2009)]{2009ApJ...696L.115F} Furukawa, N., Dawson, J.~R., Ohama, A., et al.\ 2009, \apjl, 696, L115 
\bibitem[Giannini et al.(2012)]{2012A&A...539A.156G} Giannini, T., Elia, D., Lorenzetti, D., et al.\ 2012, \aap, 539, A156 
\bibitem[Habe \& Ohta(1992)]{1992PASJ...44..203H} Habe, A., \& Ohta, K.\ 1992, \pasj, 44, 203 
\bibitem[Hill et al.(2011)]{2011A&A...533A..94H} Hill, T., Motte, F., Didelon, P., et al.\ 2011, \aap, 533, A94 
\bibitem[Hunt-Cunningham et al.(2002)]{2002aprm.conf..145H} Hunt-Cunningham, M.~R., Whiteoak, J.~B., \& Priestley, P.\ 2002, 8th Asian-Pacific Regional Meeting, Volume II, 145 
\bibitem[Inoue \& Fukui(2013)]{2013ApJ...774L..31I} Inoue, T., \& Fukui, Y.\ 2013, \apjl, 774, L31 
\bibitem[Liseau et al.(1992)]{1992A&A...265..577L} Liseau, R., Lorenzetti, D., Nisini, B., Spinoglio, L., \& Moneti, A.\ 1992, \aap, 265, 577 
\bibitem[Ladd et al.(2005)]{2005PASA...22...62L} Ladd, N., Purcell, C., Wong, T., \& Robertson, S.\ 2005, PASA, 22, 62 
\bibitem[Lo et al.(2014)]{2014ApJ...797L..17L} Lo, N., Cunningham, M.~R., Jones, P.~A., et al.\ 2014, \apjl, 797, L17 
\bibitem[Martins et al.(2010)]{2010A&A...510A..32M} Martins, F., Pomar{\`e}s, M., Deharveng, L., Zavagno, A., \& Bouret, J.~C.\ 2010, \aap, 510, A32 
\bibitem[Massi et al.(2003)]{2003A&A...399..147M} Massi, F., Lorenzetti, D., \& Giannini, T.\ 2003, \aap, 399, 147 
\bibitem[May et al.(1988)]{1988A&AS...73...51M} May, J., Murphy, D.~C., \& Thaddeus, P.\ 1988, \aaps, 73, 51 
\bibitem[Moriguchi et al.(2001)]{2001PASJ...53.1025M} Moriguchi, Y., Yamaguchi, N., Onishi, T., Mizuno, A., \& Fukui, Y.\ 2001, \pasj, 53, 1025 
\bibitem[Muench et al.(2008)]{2008hsf1.book..483M} Muench, A., Getman, K., Hillenbrand, L., \& Preibisch, T.\ 2008, Handbook of Star Forming Regions, Volume I, 4, 483 
\bibitem[Murphy \& May(1991)]{1991A&A...247..202M} Murphy, D.~C., \& May, J.\ 1991, \aap, 247, 202 
\bibitem[Nishimura et al.(2015)]{2015ApJS..216...18N} Nishimura, A., Tokuda, K., Kimura, K., et al.\ 2015, \apjs, 216, 18 
\bibitem[Ohama et al.(2010)]{2010ApJ...709..975O} Ohama, A., Dawson, J.~R., Furukawa, N., et al.\ 2010, \apj, 709, 975 
\bibitem[Oka et al.(2012)]{2012ApJS..201...14O} Oka, T., Onodera, Y., Nagai, M., et al.\ 2012, \apjs, 201, 14
\bibitem[Okamoto et al.(2017)]{2017ApJ...838..132O} Okamoto, R., Yamamoto, H., Tachihara, K., et al.\ 2017, \apj, 838, 132 
\bibitem[Oliveira(2008)]{2008hsf2.book..599O} Oliveira, J.~M.\ 2008, Handbook of Star Forming Regions, Volume II, 5, 599 
\bibitem[Ridge et al.(2006)]{2006AJ....131.2921R} Ridge, N.~A., Di Francesco, J., Kirk, H., et al.\ 2006, \aj, 131, 2921 
\bibitem[Saigo et al.(2017)]{2017ApJ...835..108S} Saigo, K., Onishi, T., Nayak, O., et al.\ 2017, \apj, 835, 108 
\bibitem[Sorai et al.(2000)]{2000SPIE.4015...86S} Sorai, K., Sunada, K., Okumura, S.~K., et al.\ 2000, \procspie, 4015, 86 
\bibitem[Takahira et al.(2014)]{2014ApJ...792...63T} Takahira, K., Tasker, E.~J., \& Habe, A.\ 2014, \apj, 792, 63 
\bibitem[Torii et al.(2011)]{2011ApJ...738...46T} Torii, K., Enokiya, R., Sano, H., et al.\ 2011, \apj, 738, 46 
\bibitem[Torii et al.(2015)]{2015ApJ...806....7T} Torii, K., Hasegawa, K., Hattori, Y., et al.\ 2015, \apj, 806, 7 
\bibitem[Vallee(1995)]{1995AJ....110.2256V} Vallee, J.~P.\ 1995, \aj, 110, 2256 
\bibitem[Verma et al.(1994)]{1994A&A...284..936V} Verma, R.~P., Bisht, R.~S., Ghosh, S.~K., et al.\ 1994, \aap, 284, 936 
\bibitem[Voisin et al.(2016)]{2016MNRAS.458.2813V} Voisin, F., Rowell, G., Burton, M.~G., et al.\ 2016, \mnras, 458, 2813
\bibitem[Wang et al.(1994)]{1994ApJS...95..503W} Wang, Y., Jaffe, D.~T., Graf, U.~U., \& Evans, N.~J., II 1994, \apjs, 95, 503 
\bibitem[Whiteoak \& Gardner(1977)]{1977PASAu...3..147W} Whiteoak, J.~B., \& Gardner, F.~F.\ 1977, Proceedings of the Astronomical Society of Australia, 3, 147 
\bibitem[Wouterloot \& Brand(1989)]{1989A&AS...80..149W} Wouterloot, J.~G.~A., \& Brand, J.\ 1989, \aaps, 80, 149 
\bibitem[Yamaguchi et al.(1999)]{1999PASJ...51..775Y} Yamaguchi, N., Mizuno, N., Saito, H., et al.\ 1999, \pasj, 51, 775 
\bibitem[Zinnecker \& Yorke(2007)]{2007ARA&A..45..481Z} Zinnecker, H., \& Yorke, H.~W.\ 2007, \araa, 45, 481 
\end{thebibliography}
\end{document}